\documentclass[useAMS,usenatbib,fleqn]{mn2e}  
\usepackage{mathptmx}
\usepackage{graphicx}
\usepackage{amsmath}
\usepackage{amssymb}
\usepackage{psfig}
\usepackage{multirow}
\usepackage{cases}
\usepackage{subfigure}
\usepackage{float}
\usepackage[caption = false]{subfig}

\def\Msun{\hbox{$\thinspace M_{\odot}$}}

\def\xmm{{\it XMM-Newton}}
\newcommand{\xmmn}{{\it XMM-Newton~\/}}

\def\nustar{{\it NuSTAR}}

\def\Mbh{\hbox{$M_{\rmn{BH}}$}}

\def\Mdot{\hbox{$\dot M$}}

\def\mdotedd{\hbox{$\dot m_{\rm Edd}$}}

\def\Ledd{\hbox{$L_{\rm Edd}$}}
\def\xmm{{\it XMM-Newton}}
\newcommand{\ms}{MS 2254.9--3712}
\newcommand{\rej}{RE J1034+396}
\newcommand{\pg}{PG 1244+026}
\newcommand{\grs}{GRS 1915+105}

\def\gsim{\mathrel{\hbox{\rlap{\hbox{\lower4pt\hbox{$\sim$}}}\hbox{$>$}}}}
\def\lsim{\mathrel{\hbox{\rlap{\hbox{\lower4pt\hbox{$\sim$}}}\hbox{$<$}}}}

   \title[The QPO in \ms]{Discovery of a $\sim 2$\,hr high frequency X-ray QPO and iron K$\alpha$ reverberation in the active galaxy \ms}
   \author[W. N. Alston et al.]{W. N. Alston$^{1}$\thanks{wna@ast.cam.ac.uk}, M. L. Parker$^{1}$, J. Markevi\v{c}i\={u}t\.{e}$^{2}$, A. C. Fabian$^{1}$, M. Middleton$^{1}$, \and A. Lohfink$^{1}$, E. Kara$^{1}$ and C. Pinto$^{1}$\\
        $^{1}$Institute of Astronomy, Madingley Rd, Cambridge, CB3 0HA.\\
	$^{2}$Department of Applied Mathematics and Theoretical Physics, Centre for Mathematical Sciences, Wilberforce Rd, Cambridge CB3 0WA.
}

\date{Accepted 2015 February 13.  Received 2015 February 4; in original form 2014 November 3}
 
\pagerange{\pageref{firstpage}--\pageref{lastpage}} \pubyear{2014}

\begin{document}
\label{firstpage}
\maketitle

\begin{abstract}
We report the discovery of a $\sim 1.5 \times 10^{-4}$\,Hz ($\sim 2$\,hr) X-ray quasi-periodic oscillation (QPO) in the active galaxy \ms, using a $\sim 70$\,ks \xmmn observation.  The QPO is significantly detected ($\sim 3.3 \sigma$) in the $1.2 - 5.0$\,keV band only, connecting its origin with the primary X-ray power-law continuum.  We detect a highly coherent soft lag between the $0.3 - 0.7$\,keV and $1.2 - 5.0$\,keV energy bands at the QPO frequency and at a frequency band in a 3:2 ratio, strongly suggesting the presence of a QPO harmonic.  An iron K$\alpha$ reverberation lag is found at the harmonic frequency, indicating the reflecting material subtends some angle to the primary continuum, which is modulated by the QPO mechanism.  Frequency resolved spectroscopy reveals the QPO and harmonic to have a hard energy dependence.  These properties of the QPO variability, together with the current black hole mass estimate, $\Mbh \sim 4 \times 10^{6} \Msun$, are consistent with the QPO originating from the same process as the high frequency QPO phenomenon observed in black hole X-ray binaries.  Principle component analysis reveals the spectral variability in \ms~is similar to that of the active galaxy \rej, a source which also displays an X-ray QPO.  This suggests a distinct spectral variability pattern for accreting black holes when in a state where QPOs are present.


\end{abstract}


%

\section{Introduction}
\label{sect:intro}

Black hole X-ray binaries (BHBs; $\Mbh \sim 10 \Msun$) display both low frequency ($\sim 0.1 - 30$\,Hz) and high frequency ($\sim 40-450$\,Hz) quasi-periodic oscillations (LFQPOs; HFQPOs) in their X-ray power spectra (PSD; see e.g. \citealt{RemillardMcClintock06} for a review).  HFQPOs are the fastest coherent features observed in accreting black holes and their high frequencies suggest an origin in the innermost regions of the accretion flow.  Understanding this phenomenon will then provide important information on the BH mass and spin as well as the structure of the strongly-curved spacetime close to the event horizon (e.g. \citealt{MilsomTaam97}; \citealt{NowakETAL97}; \citealt{Wagoner99}; \citealt{stella99}; \citealt{AbramowiczKluzniak01}; \citealt{RezzollaETAL03}; \citealt{DasCzerny11}).

A scale invariance of the accretion process (e.g. \citealt{shaksuny73}; \citealt{mushotzky93}) implies that QPOs should also be present in active galactic nuclei (AGN; $\Mbh \gsim 10^{6} \Msun$).  For a BH mass ratio $M_{\rm BHB} / M_{\rm AGN} = 10^{-5}$ the expected frequency of LFQPOs in AGN is $f_{\rm LFQPO} \gsim 10^{-5}$ (i.e. timescales of $\gsim 100$\,ks).  LFQPOs are therefore not expected to be easily detected in AGN with existing data (see \citealt{VaughanUttley05}).  The analogous HFQPOs in AGN are expected to occur at $f_{\rm HFQPO} \gsim 5 \times 10^{-3}$\,Hz (i.e. timescales of $\gsim 200$\,s), well within the temporal passband of e.g. \xmmn.  If detected in AGN, the longer periodicity means we can study the QPO on the level of individual periods, given sufficient data quality, providing a better window into the HFQPO phenomenon across the BH mass scale.

QPOs have been notoriously difficult to detect in AGN, with many early `detections' disfavoured due to an inadequacy in modelling the underlying broad band noise (\citealt{vaughan05a}; \citealt{VaughanUttley06}; \citealt{GonzalezMartinVaughan12} and references therein).  A $\sim 200$\,s QPO (most likely a HFQPO) was detected in the tidal disruption event (TDE) Swift J164449.3+573451 (\citealt{ReisETAL13}).  Recently, a $\sim 3.8$\,hr QPO was reported in 2XMM J123103.2+110648 (\citealt{LinETAL13}).  The low black hole mass ($\Mbh \sim 10^{5} \Msun$) and 50 per cent rms variability (typically $\gsim 15$ per cent for LFQPOs in BHBs, e.g. \citealt{RemillardMcClintock06}) led the authors to associate this with the LFQPO phenomenon.  The first robust AGN HFQPO detection came from the Seyfert galaxy \rej, with a $\sim 1$\,hr periodicity (\citealt{gierlinski08}).  Recently, we showed that the QPO is present in 5 years of \xmmn observations of \rej~(\citealt[][hereafter A14]{alston14b}).  The frequency of the QPO has remained constant in this time, although it is now only detected in the 1.0--4.0\,keV band.  This strengthens the association of the QPO with the primary (hot, optically thin) Comptonising corona, as observed in BHBs (e.g. \citealt{RemillardMcClintock06}).

Accreting BHs display hard lags at low frequencies --- where variations in harder energy bands are delayed with respect to softer energy bands (\citealt{miyamoto89}; \citealt{nowak99}a; \citeyear{nowak99b}b; \citeauthor[][2003a]{vaughan03a}; \citealt{mchardy04}; \citealt{arevaloetal08}; \citealt{kara13b}; \citealt{alston14a}; \citealt{LobbanETAL14}).  The leading model for the origin of the hard lags is the radial propagation of random accretion rate fluctuations through a stratified corona (e.g. \citealt{Lyubarskii97}; \citealt{churazov01}; \citealt{kotov01}, \citealt{arevalouttley06}).

A switch from hard (\emph{propagation}) lags at lower frequencies to soft (\emph{reverberation}) lags at higher frequencies has now been observed in $\gsim 20$ AGN (e.g. \citealt{fabian09}; \citealt{emmanoulopoulos11}; \citealt{zoghbi11a}; \citealt{alston13b}; \citealt{cackett13}; \citealt{demarco13lags}; \citealt{kara13a}; \citealt{alston14a}).  A consistent picture is emerging where the soft lag represents the \emph{reverberation} signal as the primary X-ray emission is reprocessed by the inner accretion disc itself (e.g. \citealt{fabian09}, see \citealt{uttley14rev} for a review).  Further evidence for this scenario comes from the corresponding detection of high frequency iron K$\alpha$ lags (e.g. \citealt{zoghbi12a}; \citealt{zoghbi13a}; \citealt{kara13c}) and Compton hump reverberation lags above 10.0\,keV observed with \nustar~(\citealt{zoghbi14}; \citealt{kara15a}).  High frequency soft lags have also been observed in an ultra-luminous X-ray source (\citealt{HeilVaughan10}; \citealt{demarco13b}) and one hard state BHB (\citealt{uttley2011}).

The time lags of HFQPOs in BHBs have also been studied, with \citet{Cui99} finding a hard lag in the 67\,Hz QPO in GRS 1915+105.  Recently, \citet{Mendez13} performed a systematic study of HFQPO time lags in a sample of 4 BHBs.  They found a hard lag in all QPOs with the exception of the 35\,Hz QPO in GRS 1915+105 which displays a soft lag that increases with increasing energy separation.  The physical meaning of these HFQPO time lags is still uncertain.  However, they provide an extra diagnostic for identifying HFQPOs and for understanding their physical origin.

In this paper we present the significant detection of a QPO in the AGN \ms~and explore several aspects of the QPO spectral-variability.  \ms~is a nearby ($z = 0.039$; \citealt{stocke91}), X-ray bright (log~($L_{\rm X}) = 43.29~{\rm erg~s^{-1}}$; \citealt{grupe04}) and `unabsorbed' ($N_{\rm H} < 2 \times 10^{22}~{\rm cm}^{-1}$; \citealt{grupe04}) radio quiet (\citealt{shields03}) narrow line Seyfert 1 (NLS1) galaxy (FWHM(H$\beta) \sim 1500$\,km s$^{-1}$; \citealt{grupe04}). 

The central BH mass in \ms~derived from the empirical $R_{\rm BLR}-\lambda L_{\lambda}(5100 {\rm \AA})$ relation (e.g. \citealt{kaspi2000}) is $\Mbh \sim 4 \times 10^{6}$ (\citealt{grupe04}; \citealt{grupe10}).   The BH mass derived from the $\Mbh - \sigma$ relation (\citealt{tremaine02}) using $\sigma$(OIII) is estimated as $\Mbh \sim 10^{7}$ (\citealt{shields03}).  The Eddington rate estimated from $\lambda L_{\lambda}(5100 {\rm \AA})$ is $L_{\rm Bol}/\Ledd = 0.24$ (\citealt{grupe04}; \citealt{grupe10}).  However, \citet{wang03} suggest \ms~is accreting at super-Eddington rate ($\Mdot / \mdotedd > 1$). They found that super-Eddington accretion can lead to a limit relation between the BH mass and the FWHM of the broad lines, indicating that super-Eddington accretors radiate close to their Eddington luminosity, but accrete above the Eddington limit (see also \citealt{CollinKawaguchi04}).

The structure of this paper is as follows: in Section~\ref{sec:obs} we describe the observations and data reduction, in Section~\ref{sec:psd} we present the power spectral analysis and QPO identification.  We explore the time delays, frequency dependent energy spectra and principle component analysis in Section~\ref{sec:var}.  In Section~\ref{sec:disco} we discuss these results and the QPO identification.


\section{Observations and data reduction}
\label{sec:obs}

\begin{figure}
\centering
\includegraphics[width=0.44\textwidth,angle=0]{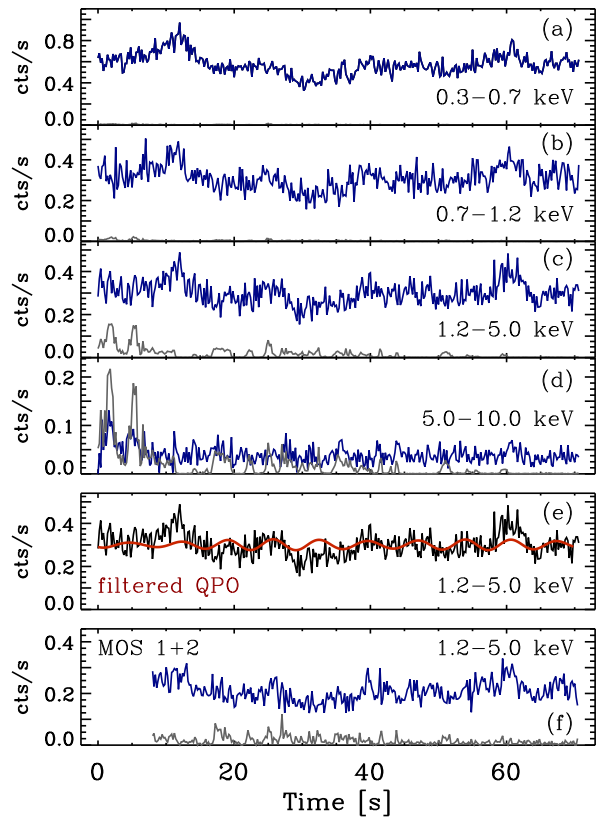}
\caption{Background-subtracted source (blue) and background (grey) EPIC-pn light curves for the 0.3--0.7\,keV (a), 0.7--1.2\,keV (b), 1.2--5.0\,keV (c) and 5.0--10.0\,keV (d) bands.  A binsize of $\Delta t = 200$\,s is used for plotting purposes.  The QPO filtered 1.2--5.0\,keV band light curve (red) and original (black) are shown in panel (e).  A bandpass filter with width $\pm 20 \%$ the QPO frequency was applied, see Section~\ref{sec:psd} for details.  Panel (f) shows the combined EPIC-MOS 1+2 background-subtracted source (blue) and background (grey) 1.2--5.0\,keV band light curves.}
\label{fig:ltcrv}
\end{figure}

We make use of the single $\sim 70$\,ks \xmmn observation of \ms~from 2005 ({\textsc OBS ID:} 0205390101).  For the timing analysis in this paper we use both the EPIC-pn (European Photon Imaging Camera; \citealt{struder01}) and EPIC-MOS data.  These observations were taken in small window mode.  The MOS data is split into 3 exposures, of which we make use of exposure 3 only with duration 62\,ks.  We processed the Observation Data Files (ODFs) following standard procedures using the \xmmn\ Science Analysis System (SAS v13.5.0), using the most recent calibration files as of October 2014.  We processed the data using the filtering conditions {\tt PATTERN} = 0--4 ($<=12$ for MOS) and {\tt FLAG} = 0 (\#XMMEA\_EM  for MOS).  We extract source light curves from a 20 arcsec circular region.  The background was taken from a large rectangular region on the same chip, approximately 15 times larger than the source region and placed away from the chip edges.  Various background regions were also used in the following analysis, and the choice of background region was found to have no significant affect on the results.

For the PSD analysis in Section~\ref{sec:psd} we filtered soft proton flares using a threshold of $0.5~{\rm ct~s}^{-1}$ in the 10.0--12.0\,keV background light curve.  We linearly interpolate across any gaps less than 500\,s and add Poisson noise.  This background rate cut removes the period of high background flaring at the start of the observation giving 58\,ks of high quality data where the EPIC-pn and EPIC-MOS overlap.  This ensures the highest signal-to-noise (S/N) light curves in the PSD analysis, which is most sensitive to uncorrelated (Poisson) noise.  After accounting for the flares at the beginning of the observation, the interpolation fraction is negligible in the remaining 58 ks.

For the remaining analysis in Section~\ref{sec:var} onwards we filter the data for soft proton flares using a threshold of $2~{\rm ct~s}^{-1}$ in the 10.0--12.0\,keV background light curve.  Again, we linearly interpolate across any short gaps and add Poisson noise, although the interpolation fraction was negligible using this rate cut.  This rate cut allows for the full 70\,ks of EPIC-pn data to be used in the cross-spectral analysis in Section \ref{sec:var}, enabling us to probe to lower frequencies and increasing the frequency resolution.

The resulting full background subtracted source light curves (with $\Delta t = 200$\,s for plotting purposes) for several energy bands are shown in Fig.~\ref{fig:ltcrv} (blue), as well as the background light curve (grey).  We show only the $1.2-5.0$\,keV MOS band for illustrative purposes.  With a mean count rate of $1.35 {\rm ~ct s^{-1}}$ in the 0.3--10.0\,keV band, pile-up is negligible in this observation.  A binsize $\Delta t = 100$\,s is used in the timing analysis throughout.


\section{Energy-dependent power spectrum}
\label{sec:psd}

\begin{figure}
  \centering
\includegraphics[width=0.3\textwidth,angle=90]{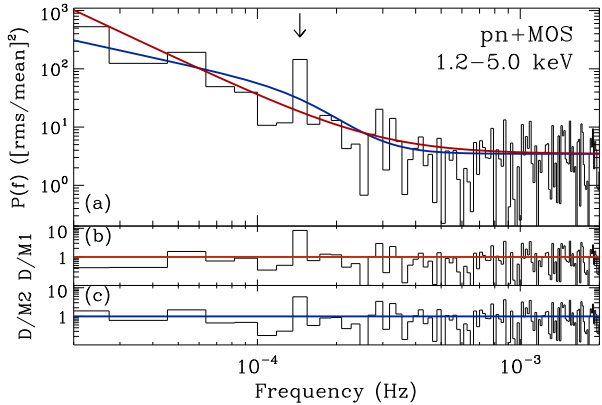}
\caption{The 1.2--5.0\,keV band PSD and model fits are shown in panel (a), for model 1 (red) and  model 2 (blue).  The data/model residuals for models 1 and 2 are shown in panels (b) and (c), respectively.}
\label{fig:psd1}
\end{figure}

The PSD was estimated using the standard method of calculating the periodogram (e.g. \citealt{priestley81}; \citealt{PercivalWalden93}), with an $\rmn{[rms/mean]}^2$ normalisation (e.g. \citeauthor[][2003a]{vaughan03a}).  Motivated by the PSD analysis in A14 we estimated the periodogram in four energy bands; 0.3--0.7\,keV, 0.7--1.2\,keV, 1.2--5.0\,keV and 5.0--10.0\,keV.  The PSD of the 1.2--5.0\,keV band is shown in Fig.~\ref{fig:psd1}.  The energy bands were chosen in order to investigate the association of a QPO feature with a particular spectral component, whilst maintaining a high signal-to-noise ratio (S/N) PSD.

\begin{figure}
  \centering
\includegraphics[width=0.3\textwidth,angle=90]{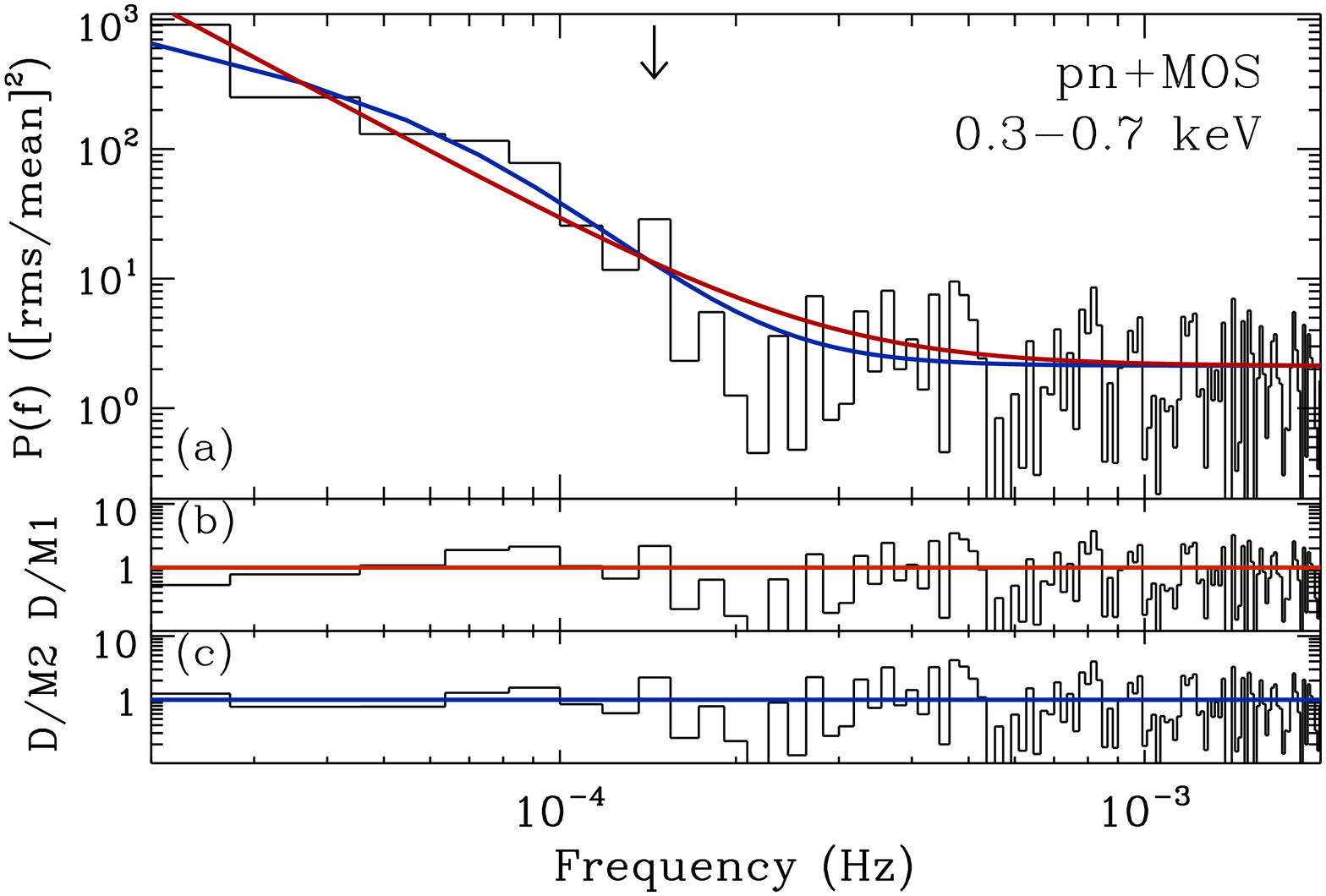}
\caption{The 0.3--0.7\,keV band PSD and model fits are shown in panel (a), for model 1 (red) and  model 2 (blue).  The data/model residuals for models 1 and 2 are shown in panels (b) and (c), respectively.}
\label{fig:psd2}
\end{figure}

\begin{figure}
  \centering
\includegraphics[width=0.3\textwidth,angle=90]{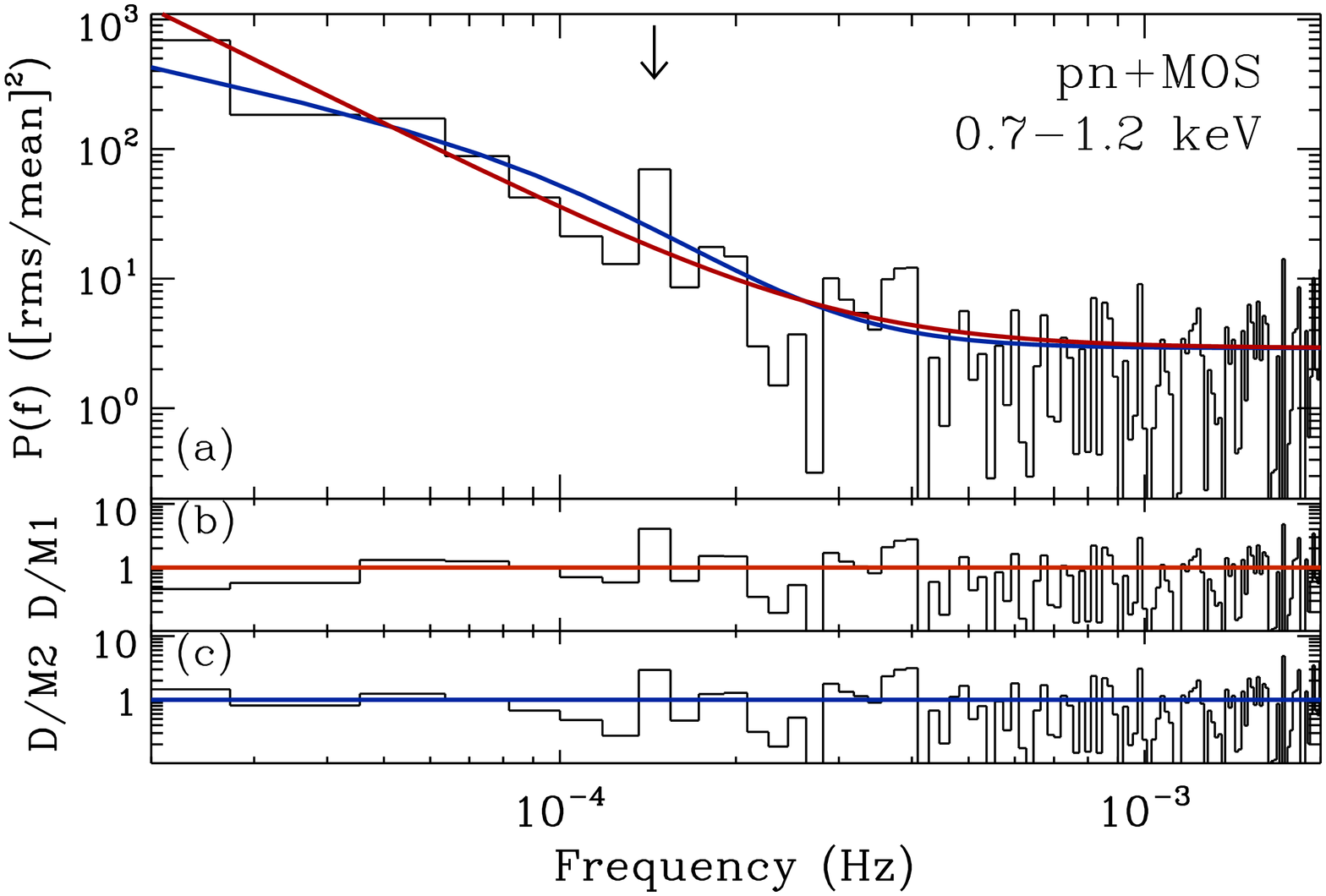}
\caption{The 0.7--1.2\,keV band PSD and model fits are shown in panel (a), for model 1 (red) and  model 2 (blue).  The data/model residuals for models 1 and 2 are shown in panels (b) and (c), respectively.}
\label{fig:psd3}
\end{figure}


\begin{table*}
\centering
\caption{Results of model fits to the pn+MOS PSDs.  Column (1) shows the energy range, column (2) shows the posterior predictive $p$-values (ppp) for the LRT between the \textit{null} hypothesis and alternative hypothesis.  Columns (3) and (4) show the ppp for $T_{\rmn{SSE}}$ and  $T_{\rmn R}$, respectively.  Column (5) shows the ratio $ R_j = 2I_j/S_j$, where $j$ is the QPO frequency, $f_{\rm QPO} = 1.5 \times 10^{-4}$\,Hz and (6) is the absolute rms at $f_{\rm QPO}$.  Columns (7), (8), (9) and (10) show the best fit model parameters with their 68.3 per cent confidence intervals.  For Model 2, column (8) shows the $\alpha_{\rm high}$ parameter.}
\begin{tabular}{l ccc ccc ccc}
\hline
{\it En} band  & $p_{\rmn{LRT}}$ & $p_{\rmn{SSE}}$  & $p_{\rmn R}$ & $R_j$ & rms($f_{\rm QPO}$) & log(N) & $\alpha$ & log($\nu_{\rm bend}$) & $P_{\rm noise}$ \\
\vspace{-0.3cm}\\
keV  &  \multicolumn{4}{c}{} & $\%$  & & & Hz&\\
(1) & (2) &(3) &(4) &(5) &(6) &(7) &(8) &(9) &(10) \\ 
\hline
  \multicolumn{10}{c}{Model 1}  \\
 0.3--0.7   & 0.1762 & 0.7750 & 0.2182  & 5  & 3 & $-9.2\substack{+1.1 \\ -1.2}$  & $2.6\substack{+0.4 \\ -0.3}$  &   -   & $0.75\substack{+0.04 \\ -0.04}$ \\
 0.7--1.2   & 0.1196 & 0.0435 & 0.0748  & 9  & 4 & $-8.5\substack{+1.3 \\ -1.5}$  & $2.4\substack{+0.4 \\ -0.3}$  &   -   & $1.07\substack{+0.05 \\ -0.05}$ \\
 1.2--5.0 * & 0.0012 & 0.0010 & 0.0010  & 18 & 6 & $-9.2\substack{+1.2 \\ -1.3}$  & $2.5\substack{+0.4 \\ -0.3}$  &   -   & $0.85\substack{+0.04 \\ -0.04}$ \\
  \multicolumn{10}{c}{Model 2}  \\
 0.3--0.7   & 0.2270 & 0.7760 & 0.7828  & 5   & 3 & $-2.2\substack{+0.3 \\ -0.4}$ & $4.6\substack{+0.4 \\ -0.7}$ & $-4.09\substack{+0.10 \\ -0.15}$ & $0.76\substack{+0.04 \\ -0.03}$ \\
 0.7--1.2   & 0.6448 & 0.0452 & 0.0757  & 6   & 4 & $-2.5\substack{+0.4 \\ -0.4}$ & $4.1\substack{+0.6 \\ -0.7}$ & $-3.92\substack{+0.11 \\ -0.15}$ & $1.06\substack{+0.04 \\ -0.05}$ \\
 1.2--5.0   & 0.5230 & 0.0252 & 0.0384  & 10  & 6 & $-2.7\substack{+0.4 \\ -0.4}$ & $4.9\substack{+0.7 \\ -0.7}$ & $-3.74\substack{+0.24 \\ -0.25}$ & $0.85\substack{+0.03 \\ -0.04}$ \\
\hline
\end{tabular}
\label{fitresults}
\end{table*}
We fitted the PSDs with simple continuum models and searched for significant data/model outliers using the maximum likelihood method of \citet[][hereafter V10]{vaughan10}.  The fitting procedure distinguishes between continuum models before testing the preferred continuum model for deficiencies that indicate the presence of a significant narrow coherent feature.  In this way were are sensitive to QPOs that are constrained to one frequency bin width.  The details of the model fitting are given in A14 and we refer the reader to V10 (and references therein) for a full discussion.  

A likelihood ratio test (LRT) statistic (eq. 22 of V10) was used to select between the continuum models, $H_0$ and $H_1$ (e.g. \citealt{ProtassovETAL02}; V10).  Following V10, the \textit{null}-hypothesis model $H_0$ was rejected using the criterion $p_{\rm LRT} < 0.01$, which the simulation results of V10 suggest is a conservative estimate.

Once the preferred continuum model has been selected, the presence of narrow coherent features is investigated using two test statistics.  Markov Chain Monte Carlo (MCMC) simulations were used to find the test statistic distribution and the associated posterior predictive $p$-value (ppp).  The overall model fit is assessed using the summed square error, $T_{\rm SSE}$ (eq. 21 of V10), which is analogous to the traditional chi-square statistic.  A small $p_{\rm SSE}$ indicates an inadequacy in the continuum modelling.  Significant outliers are investigated using $T_{\rm R} = {\rm max}_j \hat{R}_j$, where $\hat{R} = 2I_{j} / \hat{S}_j$ and $I_j$ is the observed periodogram and $S_j$ is the model power spectrum at frequency $\nu_j$.  A small $p_{\rm R}$ indicates that the largest outlier is unusual under the best-fitting continuum model and the presence of a QPO is inferred.

Following A14 (and reference therein) we use two simple continuum models; a power law plus constant (Model 1):

\begin{equation}
\label{eqn:pl}
   P(\nu) = N \nu^{- \alpha} + C
\end{equation}
\smallskip

\noindent with normalisation $N$; and the slightly more complex Model 2 is a bending power-law (e.g. \citealt{mchardy04}):

\begin{equation}
\label{eqn:bendpl}
   P(\nu) = \frac{N \nu^{{\alpha}_{\rm low}}}{1 + (\nu / \nu_{\rm bend})^{{\alpha}_{\rm low}-{{\alpha}_{\rm high}}}} + C
\end{equation}
\smallskip

\noindent where ${\alpha}_{\rm high}$ is the high frequency slope and ${\alpha}_{\rm low}$ is the slope below the bend frequency, $\nu_{\rm bend}$.  In both models, the Poisson noise level is described in the fitting process using the non-negative, additive constant, $C$.  

In model 2, $\nu_{\rm bend}$ was originally set to a value $1.5 \times 10^{-4}$\,Hz from the best fitting value found in \citet[][hereafter GMV12]{GonzalezMartinVaughan12}.  The prior distributions on all model parameters have a $3 \sigma$ dispersion around the mean level (see V10, section 9.4).  The results of the PSD fitting are shown in Table~\ref{fitresults}, along with the 68.3 per cent (1$\sigma$) confidence intervals on model parameters.


The 1.2--5.0\,keV band is the only energy band to display a significant outlier at $\sim 1.5 \times 10^{-4}$\,Hz ($p_{\rmn{SSE}} = 0.001$; $p_{\rmn R} = 0.001$), with Model 1 preferred ($p_{\rmn{LRT}} = 0.0012$).  Fig.~\ref{fig:psd1} shows the best fitting models to the 1.2--5.0\,keV band.  Despite Model 1 being preferred, the $p$-values of Model 2 are moderately low ($p_{\rmn{SSE}} = 0.03$; $p_{\rmn R} = 0.03$), indicating the outlier at $\sim 1.5 \times 10^{-4}$\,Hz is unusual under the best fitting Model 2.  In Appendix~\ref{ap:psd} we show the individual pn and MOS 1.2--5.0\,keV PSDs.



The best fitting models to the 0.3--0.7\,keV, 0.7--1.2\,keV bands are shown in Figs.~\ref{fig:psd2} and \ref{fig:psd3} respectively.  Although no formally significant outlier was detected in these two bands, some structure in the PSD can be seen at $\sim 1.5 \times 10^{-4}$\,Hz.  The 5.0--10.0\,keV band PSD is dominated by Poisson noise and hence we do not show it here.

The QPO at $\sim 1.5 \times 10^{-4}$\,Hz is confined to one frequency bin, making the QPO highly coherent.  The quality factor $Q = \nu / \Delta \nu$ is $\sim 8$.  The QPO rms fractional variability for each band is given in column 6 of Table~\ref{fitresults}.  A value of $6 \%$ is observed in the 1.2--5.0\,keV band.  If we assume the narrow outliers at $\sim 1.5 \times 10^{-4}$\,Hz in the two softer bands also indicate the presence of a QPO, then we observe an increase in the QPO rms with increasing energy band.


The QPO does not appear as apparent in the light curve compared to those observed in BHBs or the $1.0-4.0$\,keV band in \rej~(see A14 Fig.~1).  To illustrate the QPO we apply a bandpass filter to the 1.2--5.0\,keV light curve, with a frequency width $\pm 30 \%$ of the QPO frequency.  This removes the variations outside of the filter window, allowing the variations on the timescale of the filter bandpass to be seen.  The filtered light curve is plotted in Fig.~\ref{fig:ltcrv} panel (e), where the quasi-periodic nature of the light curve is now apparent.  We note here, however, that a narrow filter applied to a pure noise signal will also produce a quasi-sinusoidal time series, except the amplitude of the oscillation will be greatly reduced.  The deviation from a quasi-periodic signal in the $1.2-5.0$\,keV light curve is most likely due to the red noise nature of the broadband noise which contributes a substantial amount of power at low frequencies (see Fig.~\ref{fig:psd1}).  These long timescale trends then wash out the quasi-periodic signal.  The amplitude of the broadband noise below the observed QPO in the $1.0-4.0$\,keV band in \rej~is much smaller than in \ms, and hence why the QPO is apparent in the light curve of \rej~ (see A14 Fig.~1).


\section{The frequency dependent variability}
\label{sec:var}
\subsection{The Cross-Spectrum}
\label{sec:cspec}

\begin{figure}
  \centering
\includegraphics[width=0.4\textwidth,angle=0]{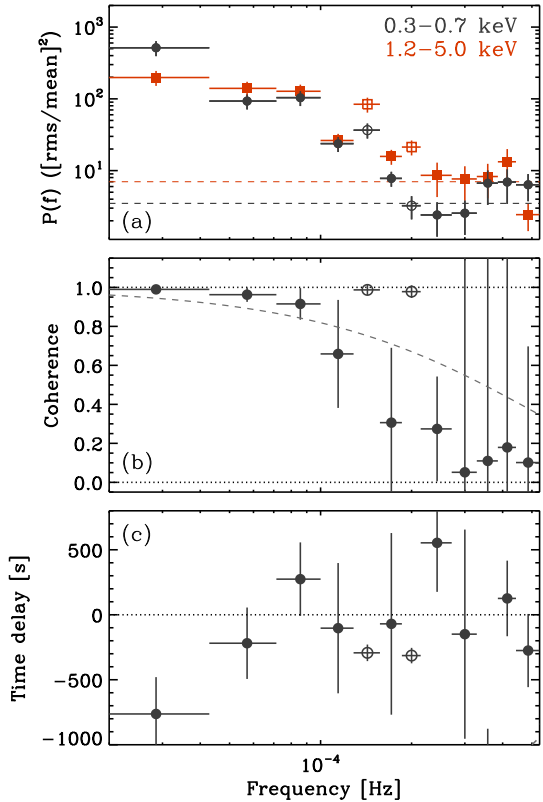}
\caption{Cross-spectral products for the $0.3-0.7$\,keV and $1.2-5.0$\,keV bands.  The data are binned as described in Sec.~\ref{sec:cspec}.  Panel (a) shows the PSDs with Poisson noise level estimates from the best fitting models described in Sec.~\ref{sec:psd}.  Panel (b) shows the Poisson-noise corrected coherence between the two bands.  The dashed line is the function $\gamma^2(f) = \exp(-f / 5 \times 10^{-4}$\,Hz).  Panel (c) shows the time delays, where a positive value indicates a hard band lag.  The open symbols highlight the estimates at the QPO and harmonic frequencies.}
\label{fig:cs}
\end{figure}

In this section we explore the cross-spectral products (PSDs, coherence and time delays) between the 1.2--5.0\,keV band and the two softer bands.  In this way we can study the frequency dependent correlations between the QPO and any components dominating at softer energies (see \citealt{uttley14rev} for a review).  In this section and in the remaining analysis we use the 70\,ks EPIC-pn observation only (see Section~\ref{sec:obs} for details).  This allows us to probe down to lower frequencies and provides more data for any segment averaging and frequency binning.

Following the method outlined in \cite{vaughannowak97} we calculate the cross-spectrum values in $M$ non-overlapping time series segments, then average over the $M$ estimates at each Fourier frequency.  To improve the signal-to-noise (S/N) in the resulting cross-spectra we averaged over neighbouring frequency bins, with each bin increasing geometrically by a factor 1.15 in frequency.  In the following analysis we use a segment length of 35\,ks and ${\rm dt} = 20$\,s.  The segment size and frequency binning are chosen in order to maximise the number of data points in each frequency bin whilst maintaining a sufficient frequency resolution to pick out any interesting features in the cross-spectral products.

Fig.~\ref{fig:cs} panel (a) shows the PSD for the $0.3-0.7$\,keV (black) and $1.2-5.0$\,keV (grey) bands.  The $1.2-5.0$\,keV shows the QPO at $\sim 1.5 \times 10^{-4}$\,Hz (open symbol).  The binned $1.2-5.0$\,keV band also shows tentative evidence for a 3:2 harmonic at $\sim 2 \times 10^{-4}$\,Hz (open symbol).  From herein we refer to this frequency as the {\it harmonic}.


From the cross-spectrum we get the coherence $\gamma^2$ (or {\it squared coherency}) between the two bands (e.g. \citealt{bendatpiersol86}).  The coherence gives a measure of the linear correlation between the two bands, i.e. how much of one band can be predicted from the other.  It is defined between [0,1], where 1 is perfectly coherent and 0 is perfectly incoherent.  Fig.~\ref{fig:cs} panel (b) shows the coherence between the $0.3-0.7$\,keV and $1.2-5.0$\,keV bands.  The coherence of the broadband noise component is high ($\sim 1$) at low frequencies and decreases with increasing frequency.  This pattern is typically observed in Seyferts (e.g. \citealt{alston13b}).  At the QPO and harmonic frequencies we see $\sim$\,unity coherence (open symbols), which sits above the falling broadband noise coherence.  Following \citeauthor[][(2003b)]{vaughan03b} we illustrate this by overlaying the function $\gamma^2(f) = \exp(-f / 5 \times 10^{-4}$\,Hz) in Fig.~\ref{fig:cs} panel (b).  This describes the coherence of the broadband noise fairly well, whilst the coherence at the QPO and harmonic frequencies are clearly distinct from this component.

From the cross-spectrum we also obtain a phase lag estimate at each frequency, $\phi(f)$, which is transformed into the corresponding time lag $\tau(f) = \phi(f) / (2 \pi f)$, with errors estimated following (\citealt{vaughannowak97}; \citealt{bendatpiersol86}).  We have previously performed extensive Monte Carlo simulations to check this method produces reliable error estimates when the contribution of Poisson noise is large (see  \citealt{alston13b}; \citealt{alston14a}).  Fig.~\ref{fig:cs} panel (c) shows the frequency dependent time lags between the $0.3-0.7$\,keV and $1.2-5.0$\,keV bands, where we follow the convention of using a positive time lag to indicate a hard band lag.  A hard lag at low frequencies is typically seen in variable Seyferts (e.g. \citealt{demarco13lags}). No significant hard lags in \ms, but do see a $\sim 3 \sigma$ negative (hereafter soft lag) at the lowest frequency.  A highly significant ($\sim 5 \sigma$) soft lag is also seen at the QPO and harmonic frequencies (open symbols).

We also measure the cross-spectral products between the $0.7-1.2$\,keV and $1.2-5.0$\,keV bands, but we do not show it here.  The coherence between the two bands is $\sim 1$ up to $\sim 3 \times 10^{-4}$\,Hz (i.e. above the harmonic frequency), above which the Poisson noise dominates.  No significant lag is found at the QPO or harmonic frequency.

\subsection{Time delays as a function of energy}

\begin{figure}
  \centering
\includegraphics[width=0.4\textwidth,angle=0]{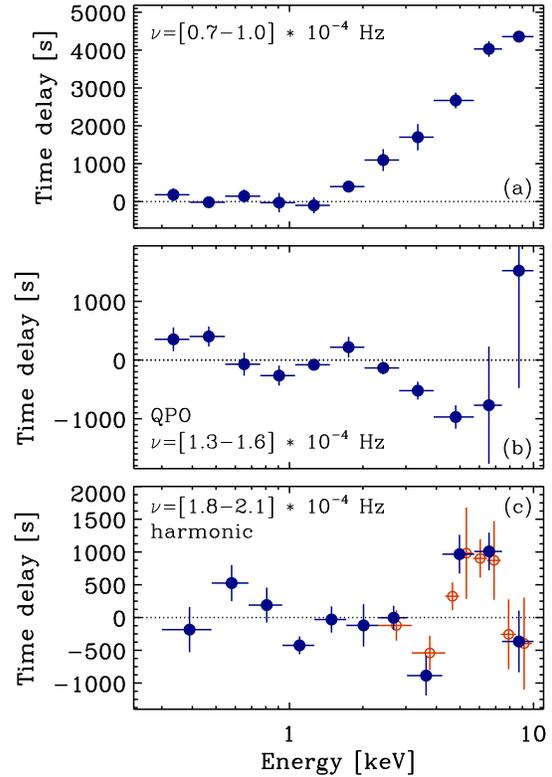}
\caption{The lag-energy spectrum for the noise ($0.7 - 1.0 \times 10^{-4}$\,Hz), QPO ($1.3 - 1.6 \times 10^{-4}$\,Hz) and harmonic ($1.8 - 2.1 \times 10^{-4}$\,Hz) frequencies.  The open circles in panel (a) are the same data but with higher energy resolution.}
\label{fig:lagen}
\end{figure}

A related technique is to study the time delays at a particular frequency as a function of energy.  The \emph{lag-energy} spectrum can be calculated by estimating the cross-spectrum between a comparison energy band vs a broad (in energy) reference band (e.g. \citealt{zoghbi11a}; \citealt{alston14a}).  If the comparison energy band falls within the reference band it is subtracted from the reference band, in order to have no correlated errors.  We use the $0.3-5.0$\,keV band as the reference band due to its high S/N.  We compute the lag-energy spectrum at a range of frequencies, including the broadband noise ($0.7 - 1.0 \times 10^{-4}$\,Hz), QPO ($1.3 - 1.6 \times 10^{-4}$\,Hz) and harmonic and ($1.8 - 2.1 \times 10^{-4}$\,Hz), and show these in Fig.~\ref{fig:lagen}.  A positive lag value indicates the average band lag compared to the broad reference band.

The broadband noise in panel (c) shows a lag that increases log-linearly between $\sim 1.0$ and 10.0\,keV.  This trend is observed at low frequencies in BHBs (e.g \citealt{miyamoto89}; \citealt{nowak99}) and AGN (e.g. \citealt{papadakis01}; \citeauthor[][2003a]{vaughan03a}; \citealt{mchardy04}; \citealt{arevaloetal08}; \citealt{kara13c}; \citealt{alston14a}; \citealt{LobbanETAL14}).  The lag-energy spectrum shows zero lag between 0.3 and $\sim 1$\,keV.

The QPO lag-energy spectrum in Fig.~\ref{fig:lagen} panel (b) shows, on average, the softer bands lagging behind the harder bands.  This is consistent with soft lag seen between the lag-frequency spectrum in Fig.~\ref{fig:cs} panel (c).  A similar lag-energy shape is observed at high frequencies in several Seyfert 1s (e.g. \citealt{kara13c}; \citealt{alston14a}).  Many of these sources also display a lag between the primary continuum (e.g. $1.0-4.0$\,keV) and the iron K$\alpha$ band at 6.4\,keV.  In the QPO lag-energy spectrum the error bars are large above $\sim 5$\,keV and no clear lag in the iron K$\alpha$ band is seen, hence we bin over from $5.0-10.0$\,keV.  This is most likely due to the low number of cross-spectral estimates being averaged in this frequency band.

The lag-energy spectrum for the harmonic frequency is shown in Fig.~\ref{fig:lagen} panel (c).  A significant ($\gsim 4 \sigma$) lag is observed between the $\sim 1 - 4$\,keV continuum band and the $\sim 5-7$\,keV band, which contains the iron K$\alpha$ band.  To show the iron K$\alpha$ lag is not sensitive to the choice of energy binning we compute the lag-energy spectrum at higher resolution (open symbols), which clearly follow the same shape.  Despite the PSD being dominated by Poisson noise above 5\,keV, we are able to pick out a significant lag at energies higher than this due to the significant power in the broad reference band and the high coherence between the energy bands. 


The average lag between the $0.3-0.7$\,keV and $1.2-5.0$\,keV bands is soft, consistent with the lag-frequency spectrum at the harmonic frequency in Fig.~\ref{fig:cs} panel (c).  A dip in the lag-energy spectrum at $3-4$\,keV compared to the remaining bands can also be seen at the harmonic in panel (c).  This feature has now been observed in the high frequency lag-energy spectra of several NLS1s (e.g. \citealt{kara13b}), but is yet to be explained.


\subsection{Frequency dependent energy spectrum}

\begin{figure}
  \centering
\includegraphics[width=0.42\textwidth,angle=0]{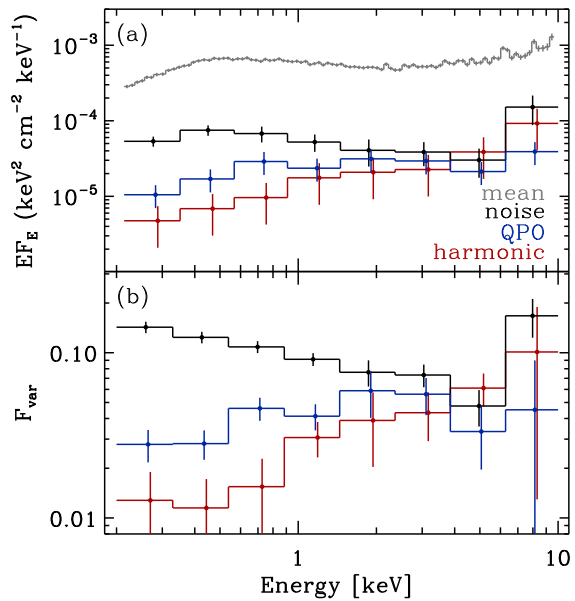}
\caption{The frequency resolved variability spectrum for the low-frequency noise ($1.4 - 4.3 \times 10^{-5}$\,Hz), QPO ($1.3 - 1.6 \times 10^{-4}$\,Hz) and harmonic ($1.9 - 2.1 \times 10^{-4}$\,Hz) frequencies. Panel (a) shows the absolute rms spectra as well as the mean (time-averaged) energy spectrum unfolded to a power law with index 0 and normalisation
1 (i.e. `fluxed' spectra).  Panel (b) shows the fractional rms spectrum ($F_{\rm var} = {\rm rms / mean}$).}
\label{fig:rms}
\end{figure}

Using frequency-resolved rms-spectra we investigate the energy dependence of the variability at different timescales (e.g.  \citealt{Edelson2002}; \citealt{MarkowitzETAL03b}; \citeauthor[][2003a]{vaughan03a}).  We calculate the rms in a given energy band by integrating the noise subtracted PSD (using an rms normalisation) over the frequency range of interest (i.e from $1/T$ to $1/2 \Delta t$).  This gives the rms spectrum in absolute units.  The fractional rms-spectrum is obtained by dividing the rms spectra by the mean count rate in each energy band.  Following \citet{PoutanenETAL08} we calculate errors using Poisson statistics.  Energy bands are made sufficiently broad such that no time bins have zero counts.

We compute the rms spectra in three frequency bands; $1.4 - 4.3 \times 10^{-5}$\,Hz (noise), $1.3 - 1.6 \times 10^{-4}$\,Hz (QPO) and $1.9 - 2.1 \times 10^{-4}$\,Hz (harmonic).  Fig.~\ref{fig:rms} shows the rms-spectra in absolute units (panel a) and fractional units (panel b).  The noise has a soft spectral shape, whereas the QPO and harmonic are both spectrally hard.  This same dependence of spectral shape with frequency is observed in the NLS1 galaxies \pg~and \rej, which are also believed to be accreting close to the Eddington rate (\citealt{MiddletonETAL09}; \citealt{middletonetal11}; \citealt{jinETAL13}).  

The time averaged spectrum is also shown in Fig.~\ref{fig:rms} panel (a).  We obtain a good fit to the data (${\Large \chi}^{2} = 773 / 784$ degrees of freedom) with a spectral model consisting of two absorbed power laws (PL) and a neutral reflection component {\sc pexmon} (\citealt{NandraETAL07}).  The spectral index of the PL components are $2.84 \pm 0.04$ and $1.43 \pm 0.06$ respectively, consistent with the values reported in \citet{BianchiETAL09}.  We use {\sc tbabs} (\citealt{wilms2000tbabs}) for the total absorption and find a value of $N_{\rm H} = 1.7 \times 10^{20}~{\rm cm}^{-2}$, consistent with the value of neutral absorption from \citet{WillingaleETAL13}.  The \xmmn RGS spectrum shows no signatures of ionised absorption.  We fit the absolute rms spectra at each frequency with a single absorbed PL, with $N_{\rm H}$ fixed as above.  The spectral index of the low-frequency noise is $2.7 \pm 0.1$, consistent with the soft component in the mean spectrum.  The spectral index of the QPO and harmonic are $2.0 \pm 0.1$ and $1.5 \pm 0.2$ respectively.  The spectral index of the harmonic is consistent with the hard PL required in the time averaged spectrum.  The smooth energy dependence of the QPO and harmonic variability suggests the QPO process is indeed present in the softer energy bands, despite the non-detection of the QPO in the PSD.

A related method for studying the variable energy spectra at a given timescale is covariance spectra (\citealt{WilkinsonUttley09}).  Using a high S/N reference band, the correlated variability is picked out in a given comparison band, thus improving the S/N of the variability spectrum compared to rms spectrum.  The covariance spectra can also be used to investigate the correlated variability between a given energy band and the remaining individual energy bands.  

We compute the covariance spectra in the Fourier domain (\citealt{uttley2011}; \citealt{CassatellaETAL2012}; \citealt{uttley14rev}) in the same frequency bands used for the rms spectra.  We use two reference energy bands to compute the covariance spectra; $0.3-0.7$ and $1.2-5.0$\, keV.  The covariance spectra from both reference bands are identical to the rms spectra shown in Fig.~\ref{fig:rms}.  The same is true for any reference band investigated.  This is unsurprising given the $\sim$~unity coherence observed between the soft and hard energy bands in Sec.~\ref{sec:cspec}.

\subsection{Principle component analysis}

\begin{figure}
  \centering
\includegraphics[width=0.42\textwidth,angle=0]{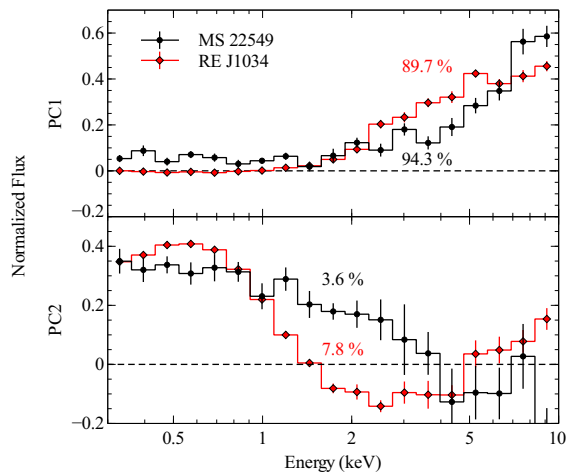}
\caption{The first (top) and second (bottom) principle components (PCs) for \ms~and \rej.  PC1 contains $\sim 90 \%$ of the variability in each source and PC2 $\sim 5 \%$.  The shape of the PCs are remarkably similar in each source and are themselves different to the two primary PCs see in other Seyferts (\citealt{parker15a}).}
\label{fig:pca}
\end{figure}

A complementary way to investigate the spectral variability is Principle Component Analysis (PCA; e.g. \citealt{VaughanFabian04}; \citealt{ParkerETAL14a}).  PCA decomposes a dataset into a set of orthogonal eigenvectors, or principal components (PCs; \citealt{Kendall75}).  When applied to X-ray data, the variability of the source spectrum is broken down into a set of variable spectral components.  If the source variability consists of a linear sum of uncorrelated and spectrally distinct physical components then an exact description of the physical components is obtained.  Whereas rms and covariance spectra are somewhat model dependent, in principle, PCA produces the individual variable spectral components in a model independent way.

Fig.~\ref{fig:pca} shows the two PCA components.  The majority of the variability is dominated by a component (PC1) that increases linearly with increasing energy band above $\sim 1$\,keV.  The second component (PC2) accounts for $\sim 5 \%$ of the variability and has a soft spectral shape.  Fig.~\ref{fig:pca} also shows the first two PCA components for \rej~(from \citealt{parker15a}).  The primary PCA components are practically identical in shape and amplitude.  The second PCA components both show a similar soft spectral dependence, however, \rej~is softer below $\sim 2.0$\,keV and is spectrally harder above $\sim 3$\,keV.  Typical NLS1s (e.g. MCG--6--30--15; \citealt{ParkerETAL14a}) have a PC1 which can be described by a soft power law (\citealt{parker15a}).  Their PC2 is due to spectral pivoting, and higher order PCs are associated with ionised reflection.  The PCA of absorption dominated Seyfert 1s also has a distinct spectral variability shape (e.g. NGC 1365; \citealt{ParkerETAL14b}).

\section{Discussion and Conclusions}
\label{sec:disco}

We have presented an analysis of the energy dependent variability in the Seyfert 1 galaxy \ms, based on a $\sim 70$\,ks \xmmn observation.  We found a significant ($\sim 3.3 \sigma$) QPO at $\sim 1.5 \times 10^{-4}$\,Hz in the PSD of the $1.2-5.0$\,keV band.  The QPO is coherent, $Q \sim 8$, and has an rms of $\sim 6$ per cent.  No significant QPO is observed in softer energy bands, although there is evidence in the PSD for some structure at the QPO frequency.

A highly coherent soft lag is seen between the $1.2-5.0$\,keV and $0.3-0.7$\,keV bands at the QPO frequency and at the frequency $\sim 2 \times 10^{-4}$\,Hz.  This strongly suggests the presence of a harmonic QPO component in a frequency ratio 3:2, although does not constitute a detection.  The coherence of the broadband noise is high at low frequencies and drops off at higher frequencies.  The highly coherent soft lag suggests that the weak periodic feature seen in the $0.3-0.7$\,keV is actually the reprocessed hard QPO emission.    

An iron K$\alpha$ lag is seen at the harmonic frequency.  If this frequency is indeed related to the QPO, then this is the first time this reverberation signature responding to a QPO modulation has been reported in the literature.   An iron K$\alpha$ reverberation lag responding to the QPO process was first observed in \rej~(\citeauthor[][{\it in prep}]{Markeviciute14prep}) making this feature unique to these two sources.

A soft lag is observed at the QPO frequency in \ms~but the data are insufficient to detect a clear iron K$\alpha$ lag at this frequency.  If it is really absent at the QPO frequency, then the lack of an iron K$\alpha$ reverberation lag suggests some geometrical dependence to the QPO and harmonic components: the disc can only respond to the harmonic oscillation, but not the QPO.

The variability of the QPO and harmonic has a hard energy dependence and are associated with the hard power law spectral component.  No significant lag is observed between the $0.7-1.2$\,keV and $1.2-5.0$\,keV bands at any frequency. This could be due to the same spectral component, modulated by the QPO process at higher frequencies, dominating the spectrum across these energies.  The X-ray and broadband energy spectrum will be investigated in detail in a follow up paper.

The similarities in the PCA of \ms~and \rej~indicates the same variability process is occurring in these two sources.  The primary PCA component in both sources has a hard spectral shape.  The QPO is also preferentially detected at harder energies, indicating the QPO has an intrinsically hard spectral shape.  \citeauthor{parker15a} measured the PCA in a sample of 26 objects and found the PCA of \rej~to be different to other well studied Seyfert 1s.  They performed extensive simulations to account for the wide range of spectral variability.  However, the hard shape of the primary PCA component in \rej~could not be reproduced.  This suggests the QPO variability is modulating the spectral components in a different way to the variability process dominating in other Seyfert 1s.


\subsection{Comparison with previous results}

GMV12 analysed the $0.2-2.0$\,keV, $2.0-10.0$\,keV and $0.2-10.0$\,keV PSD of \ms~and searched for the presence of QPOs.  They reported no significant QPO in these energy bands.  Our lack of significant detection in the  $0.3-0.7$\,keV and $0.7-1.2$\,keV bands is consistent with their results.  Our detection of a significant QPO in the $1.2-5.0$\,keV band is most likely due to the $5.0-10.0$\,keV being dominated by Poisson noise, thus affecting the detectability of the QPO over the broader energy band.  Indeed, we repeated our analysis using the energy bands of GMV12 and find no significant QPOs.

GMV12 found very weak preference for Model 2. However, their best fit parameters were very unusual for this model (e.g. $\alpha_{\rm high} \sim -8$ for the 0.2-2.0\,keV band).  Our value of $\alpha_{\rm high} \sim -4.5$ is marginally consistent with the mean slope of $\sim 3.1$ found for 15 sources with a strongly detected bend in GMV12.  Our larger value could be indicating the presence of the QPO is distorting the continuum modelling.  Our choice of higher quality data selection is most likely responsible for the different best fit model parameters found by GMV12.


\subsection{Understanding the time delays}

An approximately log-linear hard lag is observed at low frequencies ($\sim 9 \times 10^{-5}$\,Hz) where the broadband noise dominates.  This is consistent with the observed time lags at low frequencies in BHBs and AGN, which are currently best explained by the model of radial propagation of random accretion rate fluctuations through a stratified corona (e.g. \citealt{arevalouttley06}).  At the lowest frequencies investigated ($\sim 3 \times 10^{-5}$\,Hz) a tentative soft-lag is observed.  This soft lag at very low frequencies has only been observed in a handful of Seyferts: the low-flux observations of NGC 4051 (\citealt{alston13b}), MCG--6--30--15 (\citealt{kara14b}) and NGC 1365 (\citealt{kara15a}).  The origin of this low-frequency soft lag is still unclear, and whether the same mechanism is responsible in the three sources it has been detected in so far.


\citet{zoghbi11b} detected a soft lag at the QPO frequency in \rej, and found the soft lag to be broader in frequency than the QPO.  We observe the QPO and soft lag in \ms~to have the same frequency width (see Fig.~\ref{fig:cs}).  A14 found evidence for a QPO harmonic component in \rej.  The broader lag observed in \rej~may then be due to the lag at the harmonic frequency, but the current data are insufficient to individually resolve this feature.

The time lags at the HFQPO frequencies in a sample of BHBs was presented by \citet{Mendez13}.  With the exception of the 35\,Hz QPO in GRS 1915+105, all of the lags detected were hard lags, different to what we observe in \ms.  The sign of the observed lag in accreting BHs could be due to the relative fluxes of the intrinsic and lagging components (e.g. the primary power-law and soft reflection).  Indeed, \citet{alston13b} found a strong dependence on the lag direction with source flux in the NLS1 galaxy NGC 4051, with the flux changes dominated by changes in the primary continuum normalisation (e.g. \citealt{vaughan11a}).  Alternatively, the sign of the observed lag may be due to some other system parameter, such as inclination.  It is also possible that the hard lags observed by \citet{Mendez13} are in fact the same lagging process seen above 10\,keV in AGN with \nustar, which are inferred as the Compton hump lagging the primary continuum (\citealt{zoghbi14}; \citealt{kara14b}).

\subsection{QPO identification}

In this section we discuss the identification of the QPO in terms of LF or HF type.  HFQPOs in BHBs typically have a fractional rms $\sim 5 \%$ (\citealt{RemillardMcClintock06}).  The 6 per cent QPO rms in \ms~is consistent with the value observed in BHB HFQPOs, and the $\sim 8$ per cent observed in \rej~(A14).  A $Q \gsim 2$ is typically observed in HFQPOs in BHBs (e.g. \citealt{casella04}), consistent with our value of $Q \sim 8$.  HFQPOs in BHBs are observed to display harmonic components in their power spectra, with an integer ratio of 3:2 (e.g. \citealt{remillard2002, remillard2003prec}; \citealt{RemillardMcClintock06}).  We find strong evidence for the presence of a harmonic component with ratio 3:2, which strongly suggests this is a HFQPO in \ms.   

The QPO in \ms~displays many similar timing properties to the HFQPO in \rej~(\citealt{middletonetal11}) and Swift J164449.3+573451 (\citealt{ReisETAL13}).  The QPO in these two sources are also dominated by the hard X-ray component, suggesting a similar origin to the QPO in \ms.  The current best estimate of the mass of \rej~is $\Mbh \sim 1 - 4 \times 10^{6}$ (\citealt{BianHuang10}).  The factor $\sim 2$ in QPO frequency observed in these two AGN is then consistent with the factor $\sim 2$ in BH mass, if the QPO is caused by the same process which then scales linearly with black hole mass.


HFQPOs are a common feature of the very high/intermediate state (steep power-law state) in BHBs (e.g. \citealt{RemillardMcClintock06}).  These states are also characterised by mass accretion rates at or near Eddington (e.g. \citealt{Nowak1995}; \citealt{vanderklis95}).  The BHB \grs~displays HFQPOs at 35 and 67 Hz when in a super-Eddington state (e.g. \citealt{morgan97}; \citealt{Cui99}; \citealt{belloni06}; \citealt{ueda09}; \citealt{MiddletonDone10}).  \citet{wang03} have suggested that \ms~is a super-Eddington accretor.  \rej~(\citealt{middletonetal11}) and Swift J164449.3+573451 (\citealt{ReisETAL13}) are also believed to be accreting at or around Eddington.  It is then natural to associate \ms~to the high accretion rate states of BHBs, arguing in favour of a HFQPO in this source.

The HFQPO fundamental $\nu_0$ in BHBs approximately follows the relation of $\nu_0 = 931(\Mbh / \Msun )^{-1}$\,Hz, where $\nu_0$ is the unobserved QPO fundamental (\citealt{RemillardMcClintock06}).  If the QPO we detect at $\sim 1.5 \times 10^{-4}$\,Hz is the harmonic $2 \nu_0$ then we estimate $\Mbh \sim 6 \times 10^{6} \Msun$.  This is consistent with the value $\Mbh \sim 4 \times 10^{6} \Msun$ determined from the $R_{\rm BLR}-\lambda L_{\lambda}(5100 {\rm \AA})$ relation (\citealt{grupe04}).  

\citet{demarco13lags} found a close relation between the soft lag frequency and BH mass in a sample of Seyfert 1s. Assuming the same reverberation process is occurring in \ms, the frequency of the QPO soft lag would then indicate $\Mbh \sim 2 \times 10^7 \Msun$.  This is consistent with the higher BH mass estimate $\Mbh \sim 10^{7} \Msun$ (\citealt{shields03}), suggesting a HFQPO in \ms. Alternatively, if the low-frequency soft lag is caused by the same reverberation process seen in \citet{demarco13lags}, a mass of $\Mbh \sim 10^8 \Msun$ is inferred, again consistent with a HFQPO.


LFQPOs have been observed up to $\sim 30$\,Hz in BHBs (\citealt{RemillardMcClintock06}).  If the timescale of this process scales linearly with BH mass then we estimate an upper limit of $\Mbh < 1 \times 10^6 \Msun$.  Harmonic components to LFQPOs are often reported in BHBs, however the frequencies are related by the ratio 2:1.  The QPO in 2XMM J123103.2+110648 (\citealt{LinETAL13}) is only detected below $\sim 2$\,keV and no evidence for a hard PL component is seen.  The QPO rms in this source is $\sim 25 - 50$ percent, which is consistent with the rms of LFQPOs in BHBs.  When LFQPOs are present in BHBs, the broadband noise component has a flat shape, with $\alpha \sim 1$ (e.g. \citealt{RemillardMcClintock06}).  This is inconsistent with the shape of the broadband noise observed in \ms, making it unlikely to be a LFQPO.

This source currently lacks a reverberation mapping mass estimate (e.g. \citealt{Peterson04}) which is required before the exact QPO mechanism can be robustly identified.  From the arguments above we propose the QPO observed in \ms~is indeed the same as the HFQPO phenomenon observed in several BHBs.  The origin of HFQPOs is still highly uncertain, but it is clear that it must be a physical process occurring in the direct vicinity of the BH.  A long observation of \ms~is required to independently confirm the presence of the QPO and harmonic component.  If indeed the observed iron K$\alpha$ reverberation is responding to the QPO process, this will allow us to understand both these processes in better detail, and provide an important constraint for any theoretical model for the origin of HFQPOs.


\section*{Acknowledgements}

We thank the anonymous referee for constructive feedback that helped improve the manuscript.  WNA, ACF and EK acknowledge support from the European Union Seventh Framework Programme (FP7/2013--2017) under grant agreement n.312789, StrongGravity.  This paper is based on observations obtained with \xmm, an ESA science mission with instruments and contributions directly funded by ESA Member States and the USA (NASA). 

\footnotesize{
\bibliographystyle{mn2e}
\bibliography{var}
}



\appendix

\section{pn and MOS power spectra}
\label{ap:psd}

\begin{figure}
  \centering
\includegraphics[width=0.3\textwidth,angle=90]{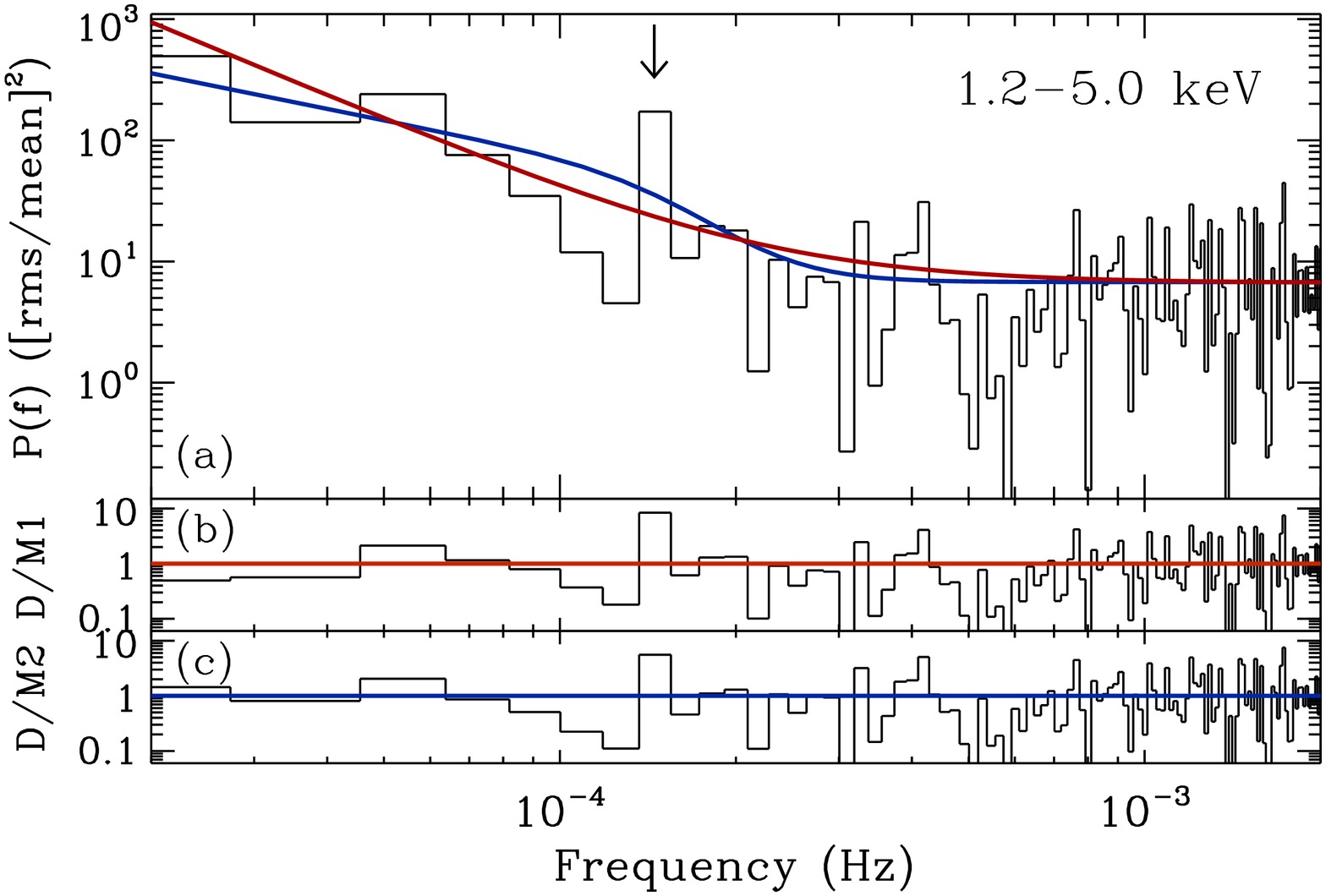}
\caption{The 1.2--5.0\,keV band PSD and model fits to the pn data are shown in panel (a), for model 1 (red) and  model 2 (blue).  The data/model residuals for models 1 and 2 are shown in panels (b) and (c), respectively.}
\label{apfig:psd1}
\end{figure}

\begin{figure}
  \centering
\includegraphics[width=0.3\textwidth,angle=90]{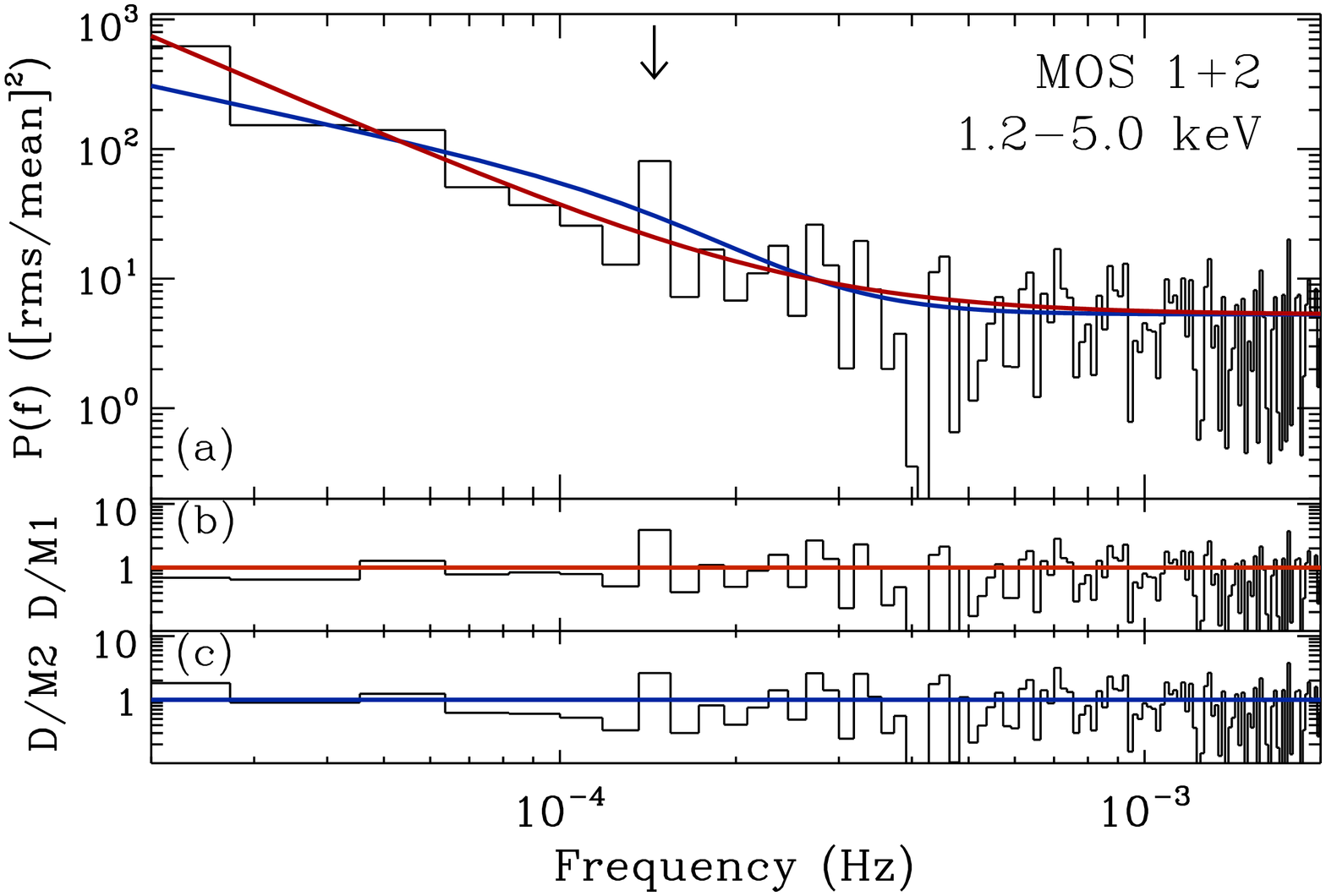}
\caption{The 1.2--5.0\,keV band PSD and model fits to the MOS data are shown in panel (a), for model 1 (red) and  model 2 (blue).  The data/model residuals for models 1 and 2 are shown in panels (b) and (c), respectively.}
\label{apfig:psd2}
\end{figure}

The PSD of \ms~presented in Sec.~\ref{sec:psd} are from the combined pn and MOS data.  Here, we show the 1.2--5.0\,keV band PSD for the individual instruments (pn; Fig.~\ref{apfig:psd1} and MOS 1+2; Fig.~\ref{apfig:psd2}).  The PSDs were computed as described in Sec.~\ref{sec:psd} using the 58\,ks `clean' light curves with $\Delta t = 100$\,s.

Both the pn and MOS 1.2--5.0\,keV band PSDs display an outlier at $\sim 1.5 \times 10^{-4}$\,Hz.  Model 1 is the preferred continuum model in both data, with $p_{\rmn{LRT}} = 0.002$ and $p_{\rmn{LRT}} = 0.024$ for the pn and MOS respectively.  For Model 1, the pn only PSD gave $p_{\rmn{SSE}} = 0.004$ and $p_{\rmn R} = 0.01$, whereas the fit to the MOS PSD gave $p_{\rmn{SSE}} = 0.028$ and $p_{\rmn R} = 0.006$.  Both these data indicate the presence of a significant outlier.  Despite Model 1 being preferred, the $p$-values of Model 2 are moderately low, with $p_{\rmn{SSE}} = 0.01$; $p_{\rmn R} = 0.04$ for the pn, and $p_{\rmn{SSE}} = 0.033$; $p_{\rmn R} = 0.005$ for the MOS.  This indicates the outlier at $\sim 1.5 \times 10^{-4}$\,Hz is also unusual under the best fitting Model 2.

\label{lastpage}


\end{document}